\documentstyle[12pt]{article}
\topmargin - 0.5in
\oddsidemargin - 5mm
\begin{document}

\def\p{\phi}
\def\P{\Phi}
\def\a{\alpha}
\def\e{\varepsilon}
\def\be{\begin{equation}}
\def\ee{\end{equation}}
\def\l{\label}
\def\0{\setcounter{equation}{0}}
\def\T{\hat{T}_}
\def\b{\beta}
\def\S{\Sigma}
\def\3{d^3{\rm \bf x}}
\def\4{d^4}
\def\C{\cite}
\def\r{\ref}
\def\ba{\begin{eqnarray}}
\def\ea{\end{eqnarray}}
\def\n{\nonumber}
\def\R{\right}
\def\L{\left}
\def\q{\hat{Q}_0}
\def\X{\Xi}
\def\x{\xi}
\def\la{\lambda}
\def\d{\delta}
\def\s{\sigma}
\def\f{\frac}
\def\vx{{\rm \bf x}}
\def\j{\frac{\delta}{i \delta j_a ({\rm \bf x},x_0+t+t_1)}}

\begin{titlepage}
\begin{flushright}
{\normalsize IP GAS-HE-7/95}
\end{flushright}

\vskip 3cm
\begin{center}
{\Large \bf The unitary transformation of the path-integral measure}
\vskip 1cm
\mbox {J.Manjavidze}
\footnote{Institute of Physics,
Georgian Academy of Science, Tamarashvili str.6,
380077, Republic of Georgia, e-mail:~jm@physics,iberiapac,ge}\\

\end{center}
\date{APRIL 1995}
\vskip 1.5cm

\begin{abstract}
\footnotesize

The aim of the article is to show how a coordinate transformation
can be applied to the path-integral formalism. For this purpose the
unitary definition of the quantum measure, which guarantees the
conservation of total probability, is  offered. As the examples, the
phase space  transformation to the canonically conjugate pare
$(energy, time)$ and the transformation to the cylindrical
coordinates are shown. The  transformations of the path-integral
measure looks classically but they can not be deduced from naive
transformations of quantum trajectories.

\end{abstract}
\end{titlepage}

\section{Introduction}
\setcounter{equation}{0}

There is the  remarkable fact that the integrable quantum systems are
quasiclassical in many aspects. For instance, the energy spectra of
the  $H$-atom problem \C {1}, of the Pocshle-Teller problem \C {2}, of the
rigid rotator problem \C {3} and even of the  $sine$-Gordon problem \C {4}
are quasiclassical. This fact should have the explanation. The Dowker's
theorem \C {5} which insist that the quasiclassical approximation is exact
for path integrals on the simple Lee group manifolds formally explains
this phenomena. So, one can expect that in curved spaces (where the
classical arguments play the crucial role) the quantum-mechanical problems
are solvable, or at least become transparent.

However we know how to construct correctly the path integral formalism
only in the  flat space \C {6}. Then in the curved space the path integrals
can be  defined through the corresponding coordinate transformations.
But there is the opinion that it is impossible to perform the
transformation of path-integral variables: the naive coordinate
transformation give wrong result since the stochastic  nature of  quantum
trajectories. One can find the examples in \C {6,7}. We intend here to
consider this problem.

The mostly powerful method of coordinate transformations in the
path-integral formalism is the ``time-sliced" method \C {8}. In frame of
this method a number of quantum problems were solved \C {9}. But it
is too cumbersome for analytical manipulations and, moreover, it leads to
unwanted time-slicing corrections (see also \C {10}).

The  stochastic nature of quantum trajectories suggests an idea
that it is possible to loss, or to add, contributions uncontrollably
when the transformations are performed. That is why such general
principle as  the conservation of total probability should play
important role. The  purpose of this article is to show how the
$S$-matrix unitarity condition can be  adopted in the path-
integral formalism to solve the problem of coordinate transformations
(preliminaries were given in \C {11}).

The unitarity condition for the  $S$-matrix
\be
SS^+ =S^+ S=1
\l{1}
\ee
presents the infinite set of nonlinear equalities for elements of
the $S$-matrix:
\be
iA A^* =A - A^*,
\l{2}
\ee
where $A$ is the amplitude, $S=1+iA$. Expressing the amplitude through the
path integral one can see that the left hand side of (\ref{2}) offers the
double integral and, at the same time, the right hand  side is the
linear combination of single integrals. Let us consider what it give to us.

Using the spectral representation of one-particle amplitude:
\be
A(x_1 ,x_2 ;E)
=\sum_{n} \Psi_{n} (x_2)
\frac{1}{E-E_n +i\varepsilon}
\Psi^*_n (x_1)
,\;\;\;\varepsilon \rightarrow +0,
\l{3}
\ee
let us calculate
\be
R(E)=\int dx_1 dx_2
A(x_1 ,x_2 ;E)A^* (x_1 ,x_2 ;E).
\l{4}
\ee
The integration over end points $x_1$ and  $x_2$ is performed for the sake
of  simplicity only. Since the orthonormalizability of the wave functions
$\Psi_{n} (x)$ we will find that
\be
R(E)=
\sum_{n}|\frac{1}{E-E_n +i\varepsilon}|^2 =
\sum_{n} \frac{i}{2\varepsilon}
(\frac{1}{E-E_n +i\varepsilon} -\frac{1}{E-E_n -i\varepsilon}) =
\frac{\pi}{\varepsilon}\sum_{n}\delta (E-E_n),
\l{5}
\ee
It is well known that
representation (\ref{3}) satisfies the  unitarity condition (\ref{2}).
But it is  important for us that $R(E)=0$ for all $E\neq E_n$. It is
evident that all unnecessary contributions with $E\neq E_n$ were
canceled by difference in the right hand  side of eq.(\ref{2}).
We will put this phenomena in the basis  of our approach.

To see the integral form of this cancelation phenomena let us use the
proper-time representation:
\be
A(x_1 ,x_2 ;E)=\sum_{n}
\Psi_{n} (x_1)\Psi^{*}_{n} (x_2)i
\int^{\infty}_{0}dTE^{i(E-E_{n}+i\varepsilon)T}
\l{6}
\ee
and insert it into (\ref{4}):
\be
R(E)=\sum_{n}
\int^{\infty}_{0}
dT_{+}dT_{-} e^{-(T_{+}+T_{-})\varepsilon}
e^{i(E-E_{n})(T_{+}-T_{-})}.
\l{7}
\ee
We  will introduce new time variables instead of $T_{\pm}$:
\be
T_{\pm}=T\pm\tau,
\l{8}
\ee
where, it follows  from Jacobian of transformation,
$|\tau|\leq T,\;\;\;0\leq T\leq \infty$. But
we can put $|\tau|\leq\infty$ since $T\sim1/\varepsilon\rightarrow\infty$
is  essential in (\ref{7}). In result,
\be
R(E)=2\pi\sum_{n}\int^{\infty}_{0}
dT e^{-2\varepsilon T}
\int^{+\infty}_{-\infty}\frac{d\tau}{\pi}
e^{2i(E-E_{n})\tau}.
\l{9}
\ee
In the last integral all contributions with $E\neq E_{n}$ are
really canceled and
\be
R(E)=\frac{\pi}{\varepsilon}\sum_{n}\delta (E-E_{n}).
\l{10}
\ee
Note that the  product of amplitudes $AA^*$ was ``linearised" after
introduction of ``virtual" time $\tau =(T_{+}-T_{-})/2$. The
physical meaning of such variables was discussed firstly in \C {12}
(see also \C {11} and Sec.3).

We will see that this cancelation mechanism unambiguously determines
the path-integral measure and, in result, it allows to perform
the transformation to arbitrary useful variables. So, in this
article we will build the perturbation theory in a curved space
for $R(E)$ omitting calculation of amplitudes. It leads to losses
of some information since the amplitudes can be restored in this
approach with a  phase accuracy only. But it is sufficient for
calculation of the energy spectrum.

In this paper the following statements will be demonstrated.

1. The unitarity condition unambiguously determines contributions
in the  path integrals.

This statement looks like a  tautology since $\exp\{iS(x)\}$,
where $S(x)$ is the action, is the unitary operator which shifts
the  system along the trajectory $x$. I.e. the unitarity is already
fixed in the path-integral formalism . But the above considered
cancellation of the real part of the  amplitude requires that the
general path-integral solution contains  the unnecessary
(i.e. unobservable) degrees of freedom. This means  that
(\ref{1}) is the necessary condition. We want to show that it is
the sufficient also,  unambiguously determines the quantum trajectories.

We start the consideration from simplest ``0-dimensional"
$x^3$ model to demonstrate quantitatively the main ideas and
technical tricks (Sec.2) and will extend the formalism on
quantum mechanics in Sec.3. The secondary result of our
approach is the demonstration of the fact that the
stationary phase method of calculation of path integrals
conserves the total probability.

2. The description of quantum-mechanical perturbations
may be reduced by the canonical transformations to the counting
of local fluctuations, with known weight function, of group
manifold.

The proof of this statement will be given in Sec.4. It based on the unitary
definition of quantum measure. The perturbation theory becomes free from
the doubling of degrees of freedom in spite of the double path
integrals are calculated.

3. The  Jacobian of transformations can be reduced to one by the
suitable chosen quantum measure, i.e. of the weight function.

We will show this in Sec.5 considering noncanonical transformation to
the cylindrical coordinates. This statement noticeably simplify the
calculations.

It must be noted that these results can not be deduced by
the naive transformations of path integral variables.

It will be  shown that even on the simple  Lee group manifolds the
quasiclassical approximation is not free from quantum corrections
in our approach, since we start the calculations from path integrals
defined in the Cartesian coordinates. The connection with the Dowker's
theorem \C{5} will be discussed briefly in Sec.6.

\section{0-dimensional model}
\setcounter{equation}{0}

Let us consider the integral:
\be
A=
\int^{+\infty}_{-\infty}
\frac{dx}{(2\pi)^{1/2}}
e^{i(\frac{1}{2}ax^2 +\frac{1}{3}bx^3)},
\l{11}
\ee
with $Im a \rightarrow +0$ and $b>0$. We want to compute the ``probability"
\be
R=|A|^2=
\int^{+\infty}_{-\infty}\frac{dx_+ dx_-}{2\pi}
e^{i(\frac{1}{2}ax_+^2 +\frac{1}{3}bx_+^3)
-i(\frac{1}{2}ax_-^2 +\frac{1}{3}bx_-^3)}.
\l{12}
\ee
As in (\ref{7}) we will introduce new variables:
\be
x_{\pm} =x\pm e.
\l{13}
\ee
In result:
\be
R=
\int^{+\infty}_{-\infty}\frac{dx de}{\pi}
e^{-2(x^2 +e^2 )Im a}e^{i(Re a\;x +2bx^2)e} e^{2i\frac{b}{3}e^3}.
\l{14}
\ee
Note that integration is  performed along the real axis for simplicity.

We  will compute the  integral over $e$ perturbatively. For this
purpose the transformation:
\be
F(e)=
e^{\frac{1}{2i}\hat{j}\hat{e}'}
e^{2ije}F(e'),
\l{15}
\ee
which is valid for any differentiable  function, is useful.
In (\ref{15}) two auxiliary variables $j$ and $e'$ were
introduced and the ``hat" symbol means the differentiation
over corresponding quantity:
\be
\hat{j}=\frac{\partial}{\partial j},\;\;\;
\hat{e'}=\frac{\partial}{\partial e'}.
\l{i}
\ee
At the end of calculations the  auxiliary variables must be
taken equal to zero.

Choosing
\be
\ln F(e)=
-2e^2 Im a +2i\frac{b}{3}e^3
\l{16}
\ee
we will find:
\be
R=
e^{\frac{1}{2i}\hat{j}\hat{e}}
\int^{+\infty}_{-\infty} dx
e^{-2(x^2 +e^2 )Im a}e^{2i\frac{b}{3}e^3}
\delta (Rea\;x +bx^2+j) .
\l{17}
\ee
Therefore, the destructive interference among two exponents in product
$AA^*$ unambiguously determines both integrals, over $x$ and over $e$.
The integral over difference $e=(x_+ -x_-)/2$ gives $\delta$-function and
then this $\delta$-function defines  the contributions in the  last
integral over $x=(x_+ +x_-)/2$. Since  $\delta$-function is the
``zero-width function" only strict solutions of equation
\be
Rea\;x +bx^2+j=0
\l{18}
\ee
give the contribution in $R$. It is the reason why the discussed
cancelation mechanism will determine unambiguously the quantum measure.

But one can note that this is not the complete solution
of the  problem: the expansion of operator exponent
$\exp \{\frac{1}{2i}\hat{j}\hat{e}\}$ generates the asymptotic
series. Note also that it is impossible to remove the
source $j$ dependence (only harmonic case $b=0$ is free from $j$).

The exponent in (\ref{11}) has two extremums at $x_1 =0$ and at
$x_2 =-a/b$. Performing trivial transformation $e\rightarrow ie$,
$\hat{e}\rightarrow -i\hat{e}$ of  auxiliary variable we  find at the limit
$Im a=0$ that the contribution from $x_1$ extremum (minimum)
gives expression:
\be
R=\frac{1}{a}
e^{-\frac{1}{2}\hat{j}\hat{e}}
(1-4bj/a^2)^{-1/2}
e^{2\frac{b}{3}e^3}
\l{19}
\ee
and the expansion of operator exponent gives  the asymptotic  series:
\be
R=\frac{1}{a}
\sum^{\infty}_{n=0}(-1)^{n}
\frac{(6n-1)!!}{n!}
(\frac{2b^4}{3a^6})^n,\;\;\;\;(-1)!!=0!!=1.
\l{20}
\ee
This series is convergent in Borel's sense \C{13}.

Let us calculate now $R$ using stationary phase method. The contribution
from the minimum $x_1$ gives $(Im a=0)$:
\be
A=e^{-i\hat{j}\hat{x}}
e^{-\frac{i}{2a}j^2}
e^{i\frac{b}{3}x^3}
(\frac{i}{a})^{1/2}.
\l{21}
\ee
The corresponding ``probability" is
\be
R=\frac{1}{a}
e^{-i(\hat{j}_+\hat{x}_+ -\hat{j}_-\hat{x}_-)}
e^{-\frac{i}{2a}(j_+^2 -j_-^2)}
e^{i\frac{b}{3}(x_+^3 -x_-^3)}.
\l{22}
\ee
Introducing the new  auxiliary variables:
\be
j_{\pm}=j \pm j_1 ,\;\;\;\;x_{\pm}= x \pm e
\l{23}
\ee
and, correspondingly,
\be
\hat{j}_{\pm}=(\hat{j}\pm\hat{j}_1)/2,\;\;\;\;
\hat{x}_{\pm}=(\hat{x}\pm\hat{e})/2
\l{24}
\ee
we find from (\ref{22}):
\be
R=\frac{1}{a}
e^{-\frac{1}{2}\hat{j}\hat{e}}
e^{2\frac{b}{3}e^3}
e^{\frac{2b}{a^2}ej^2}
\l{25}
\ee
This expression does not coincide with (\ref{19}) but it leads to
the same asymptotic series (\ref{20}). We may conclude  that both
considered methods of calculation of $R$ are equivalent since the
Borel's regularization scheme  of asymptotic series gives the
unique result.

The difference between this two methods of  calculation is  in different
organization of perturbations. So, if $F(e)$, instead of (\ref{16}), is
chosen in the form:
\be
ln F(e)=
-2e^2 Im a +2i\frac{b}{3}e^3 +2ibx^2 e,
\l{26}
\ee
we may find (\ref{25}) strightforwardly. Therefore, our method has the
freedom in choice of (quantum) source $j$.

The transition from perturbation theory with eq.(\ref{16}) to the
theory with eq.(\ref{26}) formally looks like the following
transformation of $\delta$-function:
\be
\delta (ax+bx^2 +j)=
e^{-i\hat{j}'\hat{e}'}
e^{i(bx^2 +j)e'}
\delta (ax +j').
\l{27}
\ee
Here the transformation (\ref{15}) was  used. Inserting eq.(\ref{27})
into (\ref{17}) we easily find (\ref{25}). Performing the coordinate
transformations it is useful for analytic calculations to have quantum
sources which correspond to the new variables. Formally this  transition is
equivalent to the transformation (\ref{27}). Note that this transformation
will not lead to changing of the Borel's regularization procedure.

\section{The  unitary definition of path integral measure}
\setcounter{equation}{0}

Let us consider the one dimensional motion. The  corresponding amplitude
has the form:
\be
A(x_1 ,x_2 ;E)=
i\int^{\infty}_{0}
dT e^{iET}
\int D_{C_+}x
e^{iS_{C_+}(x)}
\delta (x_1 -x(0))\delta (x_2 -x(T)),
\l{28}
\ee
where the  action
\be
S_{C_+}(x)=\int_{C_+} dt (\frac{1}{2}\dot{x}^2 -v(x))
\l{29}
\ee
and the measure
\be
D_{C_+}x=\prod_{t\in C_+}\frac{dx(t)}{(2\pi)^{1/2}}
\l{30}
\ee
are defined on the  shifted in the upper half plane Mills' time
contour $C_+ =C_+(T)$ \C{14}:
\be
t\rightarrow t+i\varepsilon,
\;\;\;\varepsilon >0, \;\;\;0\leq t \leq T.
\l{ii}
\ee
Such definition of the time  contour guaranties the convergence
of path integral.

Inserting (\ref{28}) into (\ref{4}) we find:
\ba
R(E)=
\int^{\infty}_{0}
e^{iE(T_+ -T_-)}
\int D_{C_+}x_+  D_{C_-}x_-
\times \n \\ \times
\delta (x_+ (0) -x_- (0))\delta (x_+ (T_+ ) -x_- (T_- ))
e^{iS_{C_+ (T_+ )}(x_+ ) - iS_{C_- (T_- )}(x_- )},
\l{31}
\ea
where $C_- (T)=C^{*}_{+}(T)$ is the time contour
in the lower complex half of plane. Note that the total action in
(\ref{31})
$S_{C_+ (T_+ )}(x_+ ) - S_{C_- (T_- )}(x_- )$
describes the closed-path motion which is  reversible in time.

As in Sec.1 the new time variables
\be
T_{\pm}=T\pm\tau
\l{32}
\ee
will be used. Considering $ImE\rightarrow +0$ we can consider $T$ and
$\tau$ as the independent quantities:
\be
0\leq T \leq \infty,\;\;\; -\infty \leq \tau \leq \infty.
\l{33}
\ee
Under this prescriptions the boundary condition
$x_+ (T_+ )=x_- (T_- )$ (see (\ref{31})) has simple form:
\be
x_+ (T)=x_- (T).
\l{34}
\ee
Now we will introduce mean trajectory
$x(t)=(x_+ (t) +x_- (t))/2$
and the deviation $e(t)$ from $x(t)$:
\be
x_{\pm}(t)=x(t)\pm e(t).
\l{35}
\ee
Taking into account (\ref{34}) we will have the boundary
conditions only for $e(t)$:
\be
e(0)=e(T)=0.
\l{36}
\ee
Introducing  new variables we consider $e(t)$ and $\tau$ as the
fluctuating, virtual, quantities. We will calculate the corresponding
integrals over $e$ and $\tau$ perturbatively. In zero order over $e$ and
$\tau$, i.e.  in the  quasiclassical approximation, $x$ is the  classical
path and $T$ is the total time of classical motion. Note that one can
do surely the linear transformations (\ref{35}) in the path integrals.

Let us now extract the linear over $e$ and $\tau$ terms from the
closed-path action:
\be
S_{C_+ (T_+ )}(x_+ ) - S_{C_- (T_- )}(x_- )=
-2\tau H_{T}(x)-
\int_{C^{(+)}(T)}dt
e(\ddot{x} +v'(x))-
\tilde{H}_T (x;\tau)-V_T (x,e),
\l{37}
\ee
where
\be
C^{(+)}(T)=C_+ (T)+C_-(T)
\l{iii}
\ee
is the total-time path, $H_T$ is the Hamiltonian:
\be
2H_T (x)=-\frac{\partial}{\partial T}
(S_{C_+ (T)}(x) - S_{C_- (T)}(x)),
\l{38}
\ee
and
\be
-\tilde {H}_T (x;\tau )=
S_{C_+ (T+\tau)}(x) - S_{C_- (T-\tau )}(x)+
2\tau H_T (x),
\l{39}
\ee

\be
-V_T (x,e)=S_{C_+ (T)}(x+e)-S_{C_- (T)}(x-e)+
\int_{C^{(+)}} dt e(\ddot{x}+v'(x))
\l{40}
\ee
are the remainder terms, and $v'(x)=\partial v(x)/\partial x$.  Deriving
the decomposition (\ref{37}) the definition
\be
C_- (T)=C^* _+ (T)
\l{41}
\ee
and the boundary conditions (\ref{36}) was used.

One can find the compact form of
$\exp \{-i\tilde{H}_T (x;\tau)-iV_T (x,e)\}$
expansion over $\tau$ and $e$ using formulae ({\ref{15}):
\ba
\exp\{-i\tilde{H}_T (x;\tau)-iV_T (x,e)\}=
\exp\{\frac{1}{2i}\hat{\omega}\hat{\tau}'-
i\int_{C^{(+)} (T)}dt \hat{j}(t)\hat{e}'(t)\}
\times \n \\  \times
\exp\{ 2i\omega\tau+i\int_{C^{(+)} (T)}dt j(t)e(t) \}
\exp\{-i\tilde{H}_T (x;\tau')-iV_T (x,e')\}.
\l{42}
\ea
At the end of calculations the auxiliary variables $(\omega,\tau',j,e')$
should be taken equal to zero.

Using (\ref{37}) and (\ref{42}) we find from (\ref{31}) that
\ba
R(E)=2\pi \int^{\infty}_{0}dT
\exp\{\frac{1}{2i}\hat{\omega}\hat{\tau}-
i\int_{C^{(+)} (T)}dt \hat{j}(t)\hat{e}(t)\}
\times \n \\ \times
\int Dx \exp\{-i\tilde{H}_T (x;\tau)-iV_T (x,e)\}
\delta (E+\omega -H_T (x))\prod_t \delta (\ddot{x}+v'(x)-j).
\l{43}
\ea
The  expansion over the differential operators:
\ba
\frac{1}{2i}\hat{\omega}\hat{\tau}-i\int_{C^{(+)} (T)}dt
\hat{j}(t)\hat{e}(t)=
\n \\ =
\frac{1}{2i}\frac{\partial}{\partial\omega}\frac{\partial}{\partial \tau}
-i(\int_{C+} +\int_{C_-})dt
\frac{\delta}{\delta j(t)}\frac{\delta}{\delta e(t)}=
\n \\ =
\frac{1}{2i}(\frac{\partial}{\partial\omega}\frac{\partial}{\partial \tau}
+Re\int_{C+}dt\frac{\delta}{\delta j(t)}\frac{\delta}{\delta e(t)})
\l{44}
\ea
will generate the  perturbation series. We propose  that they are sumable
in Borel sense.

The first $\delta$-function in (\ref{43}) fixes the conservation of
energy, see also (\ref{10}):
\be
E+\omega =H_T (x)
\l{45}
\ee
where $E$ is  the observed energy, $H_T (x)$ is the energy at the mean
trajectory at the time  moment $T$ and $\omega$ is the virtual
(fluctuating) energy. The second $\delta$-function
\ba
\prod_t \delta (\ddot{x}+v'(x)-j)=
(2\pi )^2 \int \prod_{t}\frac{de(t)}{\pi}
\delta(e(0))\delta(e(T))\times
\n \\ \times
e^{-2iRe\int_{C_+}dt e(\ddot{x}+v'(x)-j)}=
\prod_{t\in C_+ (T)}
\delta (Re (\ddot{x}+v'(x)-j))\delta(Im (\ddot{x}+v'(x)-j))
\l{46}
\ea
fixes the trajectory $x(t)$.

The physical meaning of $\delta$-function (\ref{46}) is   as follows
\C{11}. We can consider $(\ddot{x}+v'(x)-j)$ as the total force  and
$e(t)$ as the virtual deviation from true trajectory $x(t)$. In classical
mechanics the virtual work must be equal to zero:
\be
(\ddot{x}+v'(x)-j)e(t)=0,
\l{47}
\ee
since  the time-reversible  motion is postulated. From this evident
dynamical principle  one can find the classical equation of
motion:
\be
\ddot{x}+v'(x)-j=0,
\l{48}
\ee
since $e(t)$ is  arbitrary.

In quantum theories the  virtual work is  not equal to zero
even if the motion is reversible in time by definition (as in our
case). But integration over $e(t)$, with boundary conditions
(ref{36}) (see also Sec.6) leads to the same  result,
see (\ref{43}). So, in quantum theories the unitarity condition
play the same role as the d'Alamber's variational principle
in classical mechanics.

In (\ref{48}) $j(t)$ describes the external quantum force. The solution
$x_j (t)$ of  this equation we will find  expanding over $j(t)$:
\be
x_j (t)=x_c (t)+\int dt_1 G(t,t_1 )j(t_1 )+...
\l{49}
\ee
This is sufficient since $j(t)$ is  the auxiliary variable.
In this decomposition $x_c (t)$ is the strict solution of unperturbate
equation:
\be
\ddot{x}+v'(x)=0
\l{50}
\ee
Note that the functional $\delta$-function in (\ref{43}) does not contain
the end-point values of time $t=0$ and $t=T$. This means that the initial
conditions to the eq.(\ref{50}) are not fixed and the integration over
them is assumed since of  definition of  $R$, see (\ref{4}).

Inserting (\ref{49}) into (\ref{48}) we find the equation for Green
function:
\be
(\partial^2 +v''(x_c))_t G(t,t')=\delta (t-t').
\l{51}
\ee
It is too hard to find the exact solution of this equation since
$x_c (t)$ is  nontrivial function of $t$.  We will see that the canonical
transformation to the conserved quantities can help to avoid this
problem, see following section.

\section{Canonical transformation}
\setcounter{equation}{0}

Let us  consider motion in the phase space. For this purpose
we will insert in (\ref{43})
\be
1=\int Dp \prod_{t} \delta (p-\dot{x}).
\l{52}
\ee
In result,
\ba
R(E)=2\pi \int^{\infty}_{0}dT
\exp\{\frac{1}{2i}\hat{\omega}\hat{\tau}
-i\int_{C^{(+)} (T)}dt \hat{j}(t)\hat{e}(t)\} \times
\n \\ \times
\int Dx Dp \exp\{-i\tilde{H}_T (x;\tau)-iV_T (x,e)\} \times
\n \\ \times
\delta (E+\omega -H_T (x))
\prod_{t}\delta(\dot{x}-\frac{\partial H_j}{\partial p})
\delta(\dot{p}+\frac{\partial H_j}{\partial x}),
\l{53}
\ea
where
\be
H_{j}=\frac{1}{2}p^2 +v(x)-jx
\l{54}
\ee
is the total Hamiltonian which is time dependent through $j(t)$.

Instead of pare $(x,p)$ we introduce new pare $(\theta ,h)$
inserting in (\ref{53})
\be
1=\int D\theta Dh
\prod_{t}\delta(h-\frac{1}{2}p^2 -v(x))
\delta(\theta -\int^{x}dx (2(h-v(x)))^{-1/2}).
\l{55}
\ee
Note that the differential measures in (\ref{53}) and (\ref{55}) are
$\delta$-like. This allows to change the order of integration and
firstly integrate over $(x,p)$. Since considered transformation is
canonical, $\{h,\theta \}=1$, we find
\ba
R(E)=2\pi \int^{\infty}_{0}dT \exp\{\frac{1}{2i}\hat{\omega}\hat{\tau}-
i\int_{C^{(+)} (T)}dt \hat{j}(t)\hat{e}(t)\} \times
\n \\ \times
\int Dh D\theta \exp\{-i\tilde{H}_T (x_c ;\tau)-iV_T (x_c ,e)\}
\times \n \\ \times
\delta (E+\omega -h(T))
\prod_{t}\delta(\dot{\theta}-\frac{\partial H_c}{\partial h})
\delta(\dot{h}+\frac{\partial H_{c}}{\partial \theta}),
\l{56}
\ea
where
\be
H_c =h-jx_c (h,\theta)
\l{57}
\ee
is the transformed Hamiltonian and $x_c (\theta,h)$ is the solution of
eq.(\ref{50}) in terms   of $h$ and $\theta$.

So, instead of eq.(\ref{48}) we must solve the equations:
\be
\dot{h}=j\frac{\partial x_c}{\partial \theta},
\;\;\;\;\;\dot{\theta}=1-j\frac{\partial x_c}{\partial h},
\l{58}
\ee
which have a simple structure. Note, that
$\partial x_c/\partial \theta$ and $\partial x_c/\partial h$
in the right hand side can be  considered as the sources.
Expanding the solutions over $j$ we will find the infinite set of
recursive equations. We will use this  property of eqs.(\ref{59})
and introduce in the perturbation theory new sources:
\be
j_h =j\frac{\partial x_c}{\partial \theta},
\;\;\;\;\;j_{\theta}=j\frac{\partial x_c}{\partial h}.
\l{59}
\ee
For this purpose we will use transformation (\ref{27}):
\be
\prod_{t}\delta (\dot{h}-j\frac{\partial x_c}{\partial \theta})=
e^{-i\int_{C^{(+)}}dt\hat{j}_h (t)\hat{e}_h (t)}
e^{i\int_{C^{(+)}}e_h j\frac{\partial x_c}{\partial \theta}}
\prod_{t}\delta (\dot{h}-j_h)
\l{60}
\ee
and
\be
\prod_{t}\delta (\dot{\theta}-1+j\frac{\partial x_c}{\partial h})=
e^{-i\int_{C^{(+)}}dt\hat{j}_{\theta} (t)\hat{e}_{\theta} (t)}
e^{-i\int_{C^{(+)}}e_{\theta} j\frac{\partial x_c}{\partial h}}
\prod_{t}\delta (\dot{\theta}-1-j_{\theta}).
\l{61}
\ee
The rescaling of source $j$ lead to the rescaling of auxiliary field $e$.
In the new perturbation theory we will have two sources $j_h$, $j_{\theta}$
and two auxiliary fields $e_h$, $e_{\theta}$. Inserting (\ref{60}),
(\ref{61}) into (\ref{56}) we find:
\ba
R(E)=2\pi \int^{\infty}_{0}dT \exp \{\frac{1}{2i}\hat{\omega}\hat{\tau}-
i\int_{C^{(+)}}dt(\hat{j}_h (t)\hat{e}_h (t)+
\hat{j}_{\theta} (t)\hat{e}_{\theta} (t)) \} \times
\n \\ \times
\int Dh D\theta \exp \{ -i\tilde{H}_T (x_c ;\tau)-iV_T (x_c ,e_c ) \}
\times \n \\ \times
\delta (E+\omega -h(T))
\prod_{t}\delta (\dot{\theta}-1-j_{\theta})
\delta (\dot{h}-j_h ),
\l{62}
\ea
where
\be
e_c =
e_h \frac{\partial x_c}{\partial \theta}
-e_0 \frac{\partial x_c}{\partial h}
\l{63}
\ee
carry the simplectic structure of Hamilton's equations of motion.

Hiding the $x_c (t)$ dependence in $e_c$ we solve the  problem of the
functional determinants, see (\ref{62}), and simplify the equation
of motion as much as  possible:
\be
\dot{h}(t)=j_h (t),\;\;\;\;\;\dot{\theta}=1+j_{\theta}(t)
\l{64}
\ee
We will use the boundary conditions
\be
h(0)=h_0 ,\;\;\;\;\theta (0)=\theta_0,
\l{65}
\ee
as the extension of boundary conditions in (\ref{31}). This lead to
the following Green function of transformed perturbation theory:
\be
g(t-t')=\Theta (t-t'),
\l{66}
\ee
with the properties of projection operator:
\ba
\int dt dt' g^2 (t-t')=\int dt dt' g(t-t'),
\n \\
\int dt dt' g(t-t') g(t'-t)=0
\l{66a}
\ea
and, at the same time,
\be
g(0)=1/2.
\l{66b}
\ee
It is important to note that $Img(t)$ is  regular on the real time
axis. This  is very simplification of the perturbation theory since
it eliminate the doubling of degrees of freedom. Here one may use
that the analitical continuation to the real time axis and the
action of the perturbation-generating operator exponent are
commuting operations  (see also Sec.2 and the definition (\ref{44}).

In result, shifting $C_+$ and $C_-$ contours on the real
time axis we find:
\ba
R(E)=2\pi \int^{\infty}_{0}dT \exp\{\frac{1}{2i}(\hat{\omega}\hat{\tau}+
\int^T_0 dt_1 dt_2 \Theta (t_1 -t_2 ) (\hat{e}_h (t_1 )\hat{h}(t_2 )+
\hat{e}_{\theta}(t_1 )\hat{\theta} (t_2 )))\}
\times \n \\ \times
\int dh_0 d\theta_0 \exp\{-i\tilde{H}_T (x_c ;\tau)-iV_T (x_c ,e_c )\}
\delta (E+\omega -h_0 +h(T)),
\l{67}
\ea
where the solutions of eqs.(\ref{64})  was used. In this
expression
$x_c (t)=x_c (h_0 -h(T), t+\theta_0 -\theta (t))$ and
$(h(t),e_h (t), \theta (t),e_{\theta}(t))$ are the auxiliary fields.

Our perturbation theory describes fluctuations of the initial data
$(h_0,\theta_0 )$ of classical trajectory $x_c$. The integral form of our
perturbation theory is more complicated than of the usual one, over a
constant of interaction \C{13}, since demands the investigation of
analytic properties of $4N$-dimensional integrals, where $2N$ is  the phase
space dimension.

Here we have the new  phenomena of reduction of  quantum fluctuation,
obliviously formulated in terms of the plane waves propagation to
describe the quantum deformations of the classical trajectory, to
the ``direct" trajectory fluctuations. The direct deformations were
accumulated in the fluctuations of the classical trajectory parameters only,
i.e. in the fluctuations of the invariant hypersurface in the phase
space on which the trajectory is defined. This statement based on
the conservation of total probability.

Taking into account the point of  view of
Faddeev-Takhtajan \C{15}, that the transformation to the
initial data  of the  inverse scattering problem is the
equivalent of the canonical transformation to the
action-angle variables, this effect should have the interesting
extension to the field
theory also.

\section{Coordinate  transformation}
\setcounter{equation}{0}

In this section the coordinate transformation of two dimensional
model with potential
\be
v=v((x^{2}_{1}+x^{2}_{2})^{1/2})
\l{69}
\ee
will be considered. Repeating calculations of previous sections,
\ba
R(E)=2\pi \int^{\infty}_{0}dT
\exp\{\frac{1}{2i}\hat{\omega}\hat{\tau}-
i\int_{C^{(+)} (T)}dt \hat{\vec{j}}(t)\hat{\vec{e}}(t)\} \times
\n \\ \times
\int D^{(2)}M(x) \exp\{-i\tilde{H}_T (x;\tau)-iV_T (x,e)\},
\l{70}
\ea
where the $\delta$-like Dirak's differential measure
\be
D^{(2)}M(x) =
\delta (E+\omega -H_T (x))\prod_t \delta^{(2)} (\ddot{x}+v'(x)-j)
d^2 x(t).
\l{71}
\ee

In the classical mechanics the problem with potential (\ref{69})
is solved in the cylindrical coordinates:
\be
x_1 =r cos\phi,\;\;\;\;\;x_2 =r sin\phi.
\l{72}
\ee
We insert in (\ref{70})
\be
1=\int Dr D\phi
\prod_{t} \delta (r-(x^{2}_{1}+x^{2}_{2})^{1/2})
\delta (\phi -arc tg \frac{x_2}{x_1}).
\l{73}
\ee
to perform the transformation.
Note that the transformation (\ref{72}) is not canonical. In result
we will find a new measure:
\be
D^{(2)}M(r, \phi ) =\delta (E+\omega -H_T (x))\prod_t dr
d\phi J(r,\phi ),
\l{74}
\ee
where the Jacobian of transformation
\be
J(r,\phi )=\int \prod d^2 x
\delta^{(2)} (\ddot{x}+v'(x)-j)
\delta (\phi -arc tg \frac{x_2}{x_1})
\delta (r-(x^{2}_{1}+x^{2}_{2})^{1/2})
\l{75}
\ee
is the product of two $\delta$-functions:
\be
J(r,\phi)=\prod_t r^2 (t)
\delta (\ddot{r}-\dot{\phi}^2 r+v'(r)-j_r)
\delta(\partial_t (\dot{\phi}r^2 )-rj_{\phi}),
\l{76}
\ee
where $v'(r)=\partial v(r)/\partial r$ and
\be
j_r =j_1 \cos\phi +j_2 \sin\phi,\;\;\;\;j_{\phi}=-j_1 \sin\phi
+j_2\cos\phi
\l{77})
\ee
are  the components of $\vec{j}$ in the cylindrical
coordinates.

It is useful to organize the perturbation theory in terms of
$j_r$ and $j_{\phi}$. For this purpose the following transformation
of  arguments of $\delta$-functions  will be used:
\be
\prod_t \delta (\ddot{r}-\dot{\phi}^2 r+v'(r)-j_r)=
e^{-i\int_{C^{(+)}}dt \hat{j}'_{r}\hat{e}_r}
e^{i\int_{C^{(+)}}dt j_{r} e_r}
\prod_t \delta (\ddot{r}-\dot{\phi}^2 r+v'(r)-j'_{r})
\l{78}
\ee
and
\be
\prod_t \delta(\partial_t (\dot{\phi}r^2 )-rj_{\phi})=
e^{-i\int_{C^{(+)}}dt \hat{j}'_{\phi}\hat{e}_{\phi}}
e^{i\int_{C^{(+)}}dt j_{\phi}r e_{\phi}}
\prod_t r(t)\delta(\partial_t (\dot{\phi}r^2 )-j'_{\phi}).
\l{79}
\ee
Here $j_{r}$ and $j_{\phi}$ was defined in (\ref{77}).
In result, we get
\ba
R(E)=2\pi \int^{\infty}_{0}dT
\exp\{\frac{1}{2i}\hat{\omega}\hat{\tau}-
i\int_{C^{(+)} (T)}dt (
\hat{j}_{r}(t)\hat{e}_{r}(t)+
\hat{j}_{\phi}(t)\hat{e}_{\phi}(t))\} \times
\n \\ \times
\int D^{(2)}M(r,\phi) \exp\{-i\tilde{H}_T (x;\tau)-iV_T (x,e_{C})\},
\l{80}
\ea
where
\ba
D^{(2)}M(r, \phi ) =\delta (E+\omega -H_T (r,\phi ))
\prod_t r^2 (t)dr(t)d\phi (t) \times
\n \\ \times
 \delta (\ddot{r}-\dot{\phi}^2 r+v'(r)-j_{r})
\delta(\partial_t (\dot{\phi}r^2 )-j_{\phi})
\l{81}
\ea
and
\be
e_{C,1} =e_r \cos\phi -re_{\phi} \sin\phi,\;\;\;\;
e_{C,2}=e_r \sin\phi +re_{\phi} \cos\phi.
\l{82}
\ee
This is the  final result. The transformation looks quite classically but
(\ref{80}) can not be deduced from naive  coordinate transformation
of initial path integral for amplitude.

Inserting
\be
1=\int Dp Dl \prod_{t}
\delta  (p-\dot{r}) \delta (l-\dot{\phi}r^2)
\l{83}
\ee
in (\ref{80}) we can introduce the motion in the phase space with
Hamiltonian
\be
H_{j}=
\frac{1}{2}p^2 +\frac{l^2}{2r^2}+v(r)-j_{r}r-j_{\phi}\phi.
\l{84}
\ee
The  Dirak's  measure becomes four dimensional:
\ba
D^{(4)}M(r, \phi ,p,l) =\delta (E+\omega -H_T (r,\phi ,p,l))
\prod_t dr(t)d\phi (t) dp(t) dl(t) \times
\n \\ \times
\delta (\dot{r}-\frac{\partial H_j}{\partial p})
\delta (\dot{\phi}-\frac{\partial H_j}{\partial l})
\delta (\dot{p}+\frac{\partial H_j}{\partial r})
\delta (\dot{l}+\frac{\partial H_j}{\partial \phi})
\l{85}
\ea
Note absence of the coefficient $r^2$ in this expression. This is
the result of special choice of transformation (\ref{79}).

Since the Hamilton's group manifolds are more rich then Lagrange ones the
measure (\ref{85}) can be  considered as the starting point of
farther transformations. One must to note that the  $(action,angle)$
variables are mostly useful \C{15}. Note also that to avoid the
technical problems with equations of motion and with functional
determinants it is useful to linearise the equation of motion hiding
nonlinear terms in the corresponding auxiliary fields.

\section{Concluding remarks}
\setcounter{equation}{0}

One can note that our description looks like the description of classical
system under the influence of random $external$ force. This gives the
possibility to perform classically the transformations of the
measure. It allows also to use high resources of classical mechanics.

As the example let us consider the $H$-atom problem. The hidden
(dynamical) symmetry \C{1} means that the unperturbate Hamiltonian of
the problem $h=h(I_1 +I_2)$, where $I_k$ are the action variables.
This allows to map the problem on the $O(4)$ sphere in the (action,
angle) phase space (this transformation is not canonical: the canonical
transformation maps the problem on the   $O(2)$x$O(2)$ torus) and
following Dowker's theorem the problem on $O(4)$ sphere must be
exactly quasiclassical. But previous experience shows that each
transformation generates the corresponding sources of quantum
excitations. We will have the same for $H$-atom problem.

The solution of this problem is as follows. The expansion of the operator
exponents, which generates the perturbation series, gives the identical to
zero contribution, living only first, quasiclassical, term. The reason of
this cancellations is the special symplectic structure of the
perturbation-generating operator exponent on $O(4)$ sphere, i.e. of the
hidden symmetry.  This solution of $H$-atom problem will be published.

The interesting possibility may arise also in connection with Kolmogorov-
Arnold-Mozer (KAM) theorem \C{15}: the system which is not strictly
integrable can show the stable motion peculiar to integrable systems.
This is the argument in favor of the idea that there may be another
mechanism of suppression of the quantum excitations.

One can note that the transformed perturbation theory describes only
the retarded quantum fluctuations since $\partial x_c / \partial h_0$
and $\partial x_c / \partial \phi_0$ was considered as the sources,
see also the definition of Green function (\ref{66}). This feature
of the theory can lead to the time irreversibility of quantum
processes and must be explained.

The starting expression (\ref{31}) describes the reversible in time
motion since total action $S_{C_+ (T_+ )}(x_+ ) - S_{C_- (T_- )}(x_- )$
is time reversible. But the unitarity condition forced us to consider
the interference picture between expanding (with phase $\exp
\{iS_{C_+ (T_+ )}(x_+ ) \}$) and converging ($\exp \{-iS_{C_- (T_- )}
(x_- ) \}$) waves. This is fixed by the boundary conditions:
$x_+ (0)=x_- (0)$ and $x_+ (T_+ )=x_- (T_- )$. Despite this fact
the quantum theory remain time reversible up to canonical
transformation to the invariant hypersurface of the constant
energy. The causal Green function $G(t-t')$ , see (\ref{51}),
describes as advanced as well retarded perturbations and the
theory contains the doubling of degrees of freedom. It means
that the theory ``keep in mind" the time reversibility.

But after the canonical transformation, using above mentioned
boundary conditions, and continuing the theory to the real
times, the memory of doubling of the degrees of freedom disappears
and the theory becomes time irreversible.

The key step in this calculations was an extraction of the classical
trajectory $x_c$ which can not be defined without definition of boundary
conditions. Just this fact introduces the time vector and the quantum
perturbations (of the transformed perturbation theory) follow this vector
(in the nontransformed perturbation theory this effect is hidden in
the impossibility to find the strict solution of eq.(\ref{51})).
Therefore, the considered irreversibility of an quantum-mechanical
motion have most probably the imaginary character since the
symmetric, one particle, boundary conditions are required.

\vspace{0.2in}
{\Large \bf Acknowledgement}
\vspace{0.2in}

I would  like to thank my colleagues from the Institute of Physics
(Tbilisi), and especially A.Ushveridze, for interesting discussions. This
work was supported in part by the  U.S. National Science Foundation.


\newpage
\begin{thebibliography}{99}

\bibitem{1}V.~Fock, {\it Zs.~Phys.,\bf 98},145(1935);
V.~Bargman, {\it Zs.Phys.,\bf99},576(1935)
\bibitem{2}G.~Pocshle and E.~Teller,{\it Zs.Phys.,\bf 83},143(1933)
\bibitem{3}I.~H.~Duru,{\it Phys.~Rev.,\bf D30},2121(1984)
\bibitem{4}R.~Dashen, B.~Hasslacher and A.~Neveu,
{\it Phys.~Rev.,\bf D10}, 3424(1975)
\bibitem{5}J.~S.~Dowker, {\it Ann.~Phys.(NY),\bf 62},361(1971)
\bibitem{6}M.~S.~Marinov, {\it Phys.~Rep.,\bf 60},1(1980)
\bibitem{7}S.~F.~Edvards and Y.~Guliaev,
{\it Proc.~Roy.~Soc.,\bf A279},229(1964);
J.~R.~Klauder, {\it Acta Phys.~Austr.~Suppl.,\bf XXII},3(1980)
\bibitem{8}R.~P.~Feynman and A.~R.~Hibbs, {\it Quantum Mechanics and
Path integrals},(McGraw-Hill,~New York,~1965)
\bibitem{9}I.~H.~Duru and H.~Kleinert, {\it Phys.~Lett.,\bf 84B},185(1979);
C.~Groshe, {\it DESY, \bf 129}(1991);
R.~Ho and A.~Inomata,{\it Phys.~Rev.~Lett.,\bf 48},231(1982)
\bibitem{10}H.~Kleinert, {\it Path Integrals in Quantum Mechanics,
Statistics and Polimer Physics}, (World Scientific, Singapore, 1989)
\bibitem{11}J.~Manjavidze, {\it Sov.~Nucl.~Phys., \bf 45}, 442(1987)
\bibitem{12}V.~Fock, {\it Vestnik LGU, \bf 16}, 57 (1959)
\bibitem{13}G.~t'Hooft, {\it The Whys of Subnuclear Physics, ed. by
Zichichi}, (Plenum, New York \& London, 1977)
\bibitem{14}R.~Mills, {\it Propagators of Many-Particles Systems},
(Gordon \& Breach, 1969)
\bibitem{15}V.~I.~Arnold, {\it Mathematical methods of Classical
Mechanics}, (Springer Verlag, New York, 1978)

\end{thebibliography}
\end{document}